\documentclass[12pt]{article}

\usepackage{newtxtext,newtxmath}
\usepackage{graphicx}
\usepackage{pdfpages}
\usepackage{xcolor}
\usepackage{lineno}

\usepackage[letterpaper,margin=1in]{geometry}

\renewenvironment{abstract}
	{\quotation}
	{\endquotation}

\date{}

\makeatletter
\renewcommand{\fnum@figure}{\textbf{Figure \thefigure}}
\renewcommand{\fnum@table}{\textbf{Table \thetable}}
\makeatother

\usepackage[numbers,super,sort&compress]{natbib}

\usepackage{url}

\def\scititle{
Context-Conditioned Generative Models Enable Subnational Refinement of Sparse Humanitarian Surveys
}
\title{\bfseries \boldmath \scititle}

\author{
	Federica~Sibilla$^{1}$,
	Vasiliki~Voukelatou$^{2\ast}$,
	Duccio~Piovani$^{2}$,\and
    Kyriacos~Koupparis$^{2}$,
    Daniela~Paolotti$^{1}$,
    Rossano Schifanella$^{3,1}$,\and
    Kyriaki~Kalimeri$^{1,4\ast}$
    \and
	\small$^{1}$ISI Foundation, Torino \& 10123, Italy.
	\small$^{2}$World Food Programme, Rome \& 00148, Italy.\and\and
    \small$^{3}$Computer Science, Università di Torino, Turin \& 10124, Italy.
    \small$^{4}$UNICEF, New York \& 10038, USA.\and
	\small$^\ast$Corresponding author. Email: vasiliki.voukelatou@wfp.com, kyriaki.kalimeri@isi.it
}
\begin{document}

\maketitle

\begin{abstract}
Data scarcity limits inference in many scientific and policy domains. Survey data are essential for decision-making, but sparse samples often fail to capture fine spatial granularities. We evaluate normalizing flows, a generative model that learns complex data distributions and can be conditioned on exogenous contextual features, in controlled data scarcity scenarios. Across eight household survey datasets spanning six low-income or middle-income countries in the humanitarian domain, we show that context-conditioned generative models can refine sub-national survey distributions under severe data scarcity, and that performance increases systematically with the richness of the conditioning information. These findings support a principle for survey data augmentation: generative models can improve sub-national estimates when the sparse sample retains sufficient support and contextual covariates encode relevant local heterogeneity. Gains in distributional fidelity do not uniformly translate into better regional rankings, and we characterise the conditions under which they do. By learning full conditional distributions rather than point estimates, the approach provides fine-grained evidence for humanitarian decision-making and resource allocation.
\end{abstract}

\section*{Introduction}

Many scientific and policy domains rely on data that are sparse, costly to collect, or unevenly distributed across space and populations. Yet decision-making often requires fine-grained estimates of distributions and aggregates at sub-national levels and across demographic groups~\cite{elbers2003microlevel,wakefield2019estimating}. Classical small-area estimation approaches address this gap by borrowing strength across areas to estimate aggregate indicators such as means, totals, or poverty rates~\cite{fay1979estimates,battese1988error}, but they do not reconstruct the full household-level distributions from which multiple indicators can subsequently be derived. {Non-parametric learning extensions of this framework, which replace linear mixed models with tree-based learners and draw on geospatial covariates, improve accuracy but retain the same aggregate targets~\cite{krennmair2022flexible,newhouse2023sae,corral2025poverty}}. This tension is especially acute in humanitarian settings, where household surveys are the primary instrument for informing operational decisions. Food security assessments such as World Food Programme's (WFP) Vulnerability Analysis and Mapping surveys guide targeting and resource allocation for hunger response~\cite{wfp_vam_food_security_analysis}; socioeconomic surveys such as the World Bank Living Standards Measurement Studies inform poverty mapping and social protection design~\cite{elbers2003microlevel}; and child welfare surveys such as UNICEF's Multiple Indicator Cluster Surveys (MICS) provide the evidence base for national policy on education, nutrition, and child protection across more than 120 countries~\cite{nga_mics_2021}. Yet data collection in these settings remains expensive, infrequent, and difficult to implement in vulnerable or unstable environments, particularly in conflict-affected areas. In these contexts, researchers and practitioners often need to draw detailed inferences from samples that are too limited to recover local heterogeneity reliably.

{These constraints are concrete in the settings studied here. In Yemen, WFP monitors food security through remote telephone interviews because face-to-face data collection is often difficult to implement in complex operational environments, and survey coverage may vary across locations. In Mozambique, post-shock food security assessments must inform the prioritisation of provinces for emergency response within days, from samples drawn only in accessible areas. In each case the operational question, which sub-national units need intervention most urgently, requires finer resolution than the sample provides.}

Recent advances in generative artificial intelligence have enabled the synthesis of realistic tabular data and the modeling of complex conditional distributions~\cite{bourou2021review,wang2024review,xu2019ctgan,wang2025ttvae,rezende2015normalizing,dinh2017density,durkan2019neural,papamakarios2021normalizing}. Diffusion models and large language models have since become the reference generators for tabular data~\cite{kotelnikov2023tabddpm,zhang2024tabsyn,borisov2023great,li2025diffusion}. This has motivated researchers to apply generative models for survey data generation. Applications have largely focused on synthetic population generation for downstream agent-based model simulations, spatial microsimulation, and privacy-preserving data release~\cite{jiang2025synthetic,liu2023goggle,lim2025llmbn,jones2026llm,johnsen2022population,tanton2012spatial}. More recently, generative models have been explored as an augmentation tool in the health domain, to enlarge the training set of downstream machine learning models when clinical data are scarce~\cite{liu2025synthetic}. This latter role in statistical inference under data scarcity remains much less established~\cite{Manousakas2023useful,abdelazim2026harnessing}. A fundamental challenge is that generative models cannot create information absent from the training sample: when the training data are non-representative, biased or incomplete, the model will reproduce and may even amplify these limitations rather than correct them~\cite{shumailov2024nature,jordon2022synthetic}.

Building on the above, previous work has mostly focused on improving the fidelity of synthetic data while preserving privacy, giving less attention to the conditions under which generative models can meaningfully contribute to inference. Many of the target outcomes measured by household surveys are structurally related to contextual characteristics of a place, including market accessibility, relative wealth, and environmental conditions~\cite{jean2016satellite, yeh2020using, chi2022microestimates, martini2022food, voukelatou2026predicting}. These contextual signals are increasingly available at fine spatial resolution from satellite imagery, mobile phone networks, and remote sensing products, independently of survey collection~\cite{burke2021satellite, rolf2021mosaiks, benassai2025unequal, huang2021commentary}. This motivates the question: under what conditions can generative models integrate exogenous contextual information with sparse survey data to serve as inference tools? This question matters because many applications require more than plausible synthetic records. They require reliable estimates of local distributions and aggregate indicators in settings where direct measurement is limited.

\begin{figure}
    \centering
    \includegraphics[width=\textwidth]{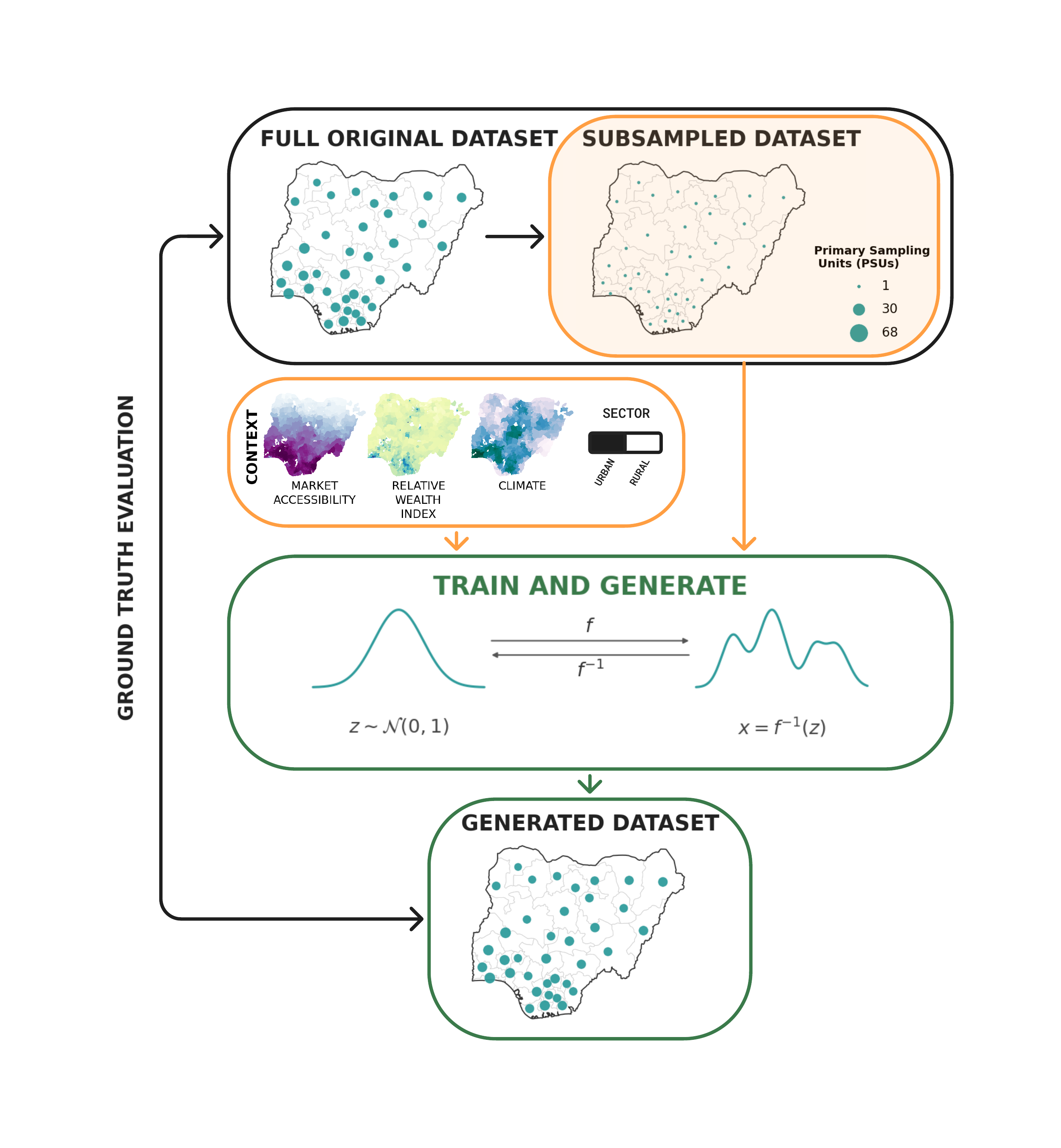}
    \caption{\textbf{Proposed framework on the illustrative example of Nigeria.} We simulate data scarcity by keeping only one survey cluster per region, train a generative AI model conditioned on contextual features, generate synthetic data for excluded locations, and compare them with held-out survey data for evaluation.}
    \label{fig:fig1}
\end{figure}

In this work, we address this question in the context of humanitarian household survey data, where sparsity, heterogeneity, and sampling constraints are pervasive. We introduce a conceptual and empirical framework based on conditional generation (Figure~\ref{fig:fig1}): a normalizing flow~\cite{winkler2019cnf} (cNF), a generative AI model, is trained on a sparse, yet nationally representative household survey subsample and conditioned on contextual features, including market accessibility~\cite{benassai2025unequal}, relative wealth indices~\cite{chi2022microestimates}, and environmental indicators~\cite{huang2021commentary}, which are available at fine spatial resolution independently of any data collection. Assuming these relationships are spatially homogeneous within a country, the model that learns how place-level signals modulate the distribution of household-level outcomes can generate synthetic household records for under-sampled (or even unsampled) locations by conditioning on their known covariate values, thereby refining local distributional structure that is only weakly represented in the sparse sample. We simulate controlled data scarcity by subsampling household survey data to deliberately disrupt sub-national representativeness while preserving national-level composition. This methodological framework is applied across eight nationally representative surveys spanning six low-income or middle-income countries: Ethiopia, Sri Lanka, Mozambique, Nigeria, Zimbabwe and Yemen were selected as priority humanitarian contexts, and comparable survey data availability enabling consistent contextual feature construction (Supplementary table 1).
We then evaluate whether the generated data better approximate sub-national distributions and
aggregate indicators than an oversampling baseline of the observed sample. We test whether generative refinement improves estimates and compare the context-conditioned model (cNF) with two versions of the normalizing flow: a model conditioned solely on a categorical geographical-region label (NF), and a model conditioned on both the geographical-region label and sector, with sector representing the rural/urban classification (NF + sector). This comparison helps isolate continuous contextual features as the source of additional generative improvement. Standard tabular generative models, including CTGAN and TVAE~\cite{xu2019ctgan} and the more recent diffusion-based generators~\cite{kotelnikov2023tabddpm,zhang2024tabsyn}, are not included in the main inferential analysis because they do not provide a natural mechanism for explicitly conditioning generation on continuous external covariates at both training and generation time, which we require to generate records for under-sampled locations. Instead, we include them in a separate supplementary benchmark, training on the full survey data, to evaluate general synthetic-data quality in terms of fidelity and originality against the full population~\cite{Sajjadi2018precisionrecall}. Across datasets, cNF achieves fidelity--originality trade-offs comparable to or better than these widely used tabular generative models (Supplementary figure 4), indicating that the observed inferential gains are not attributable to an unusually weak generative baseline.

Our results show that context-conditioned normalizing flows applied to scarce survey data, where context is represented as exogenous features linked to environmental, wealth, accessibility indicators, and settlement type,
provide more accurate, higher-resolution estimates than simple baseline models. We also explore the utility of generated data for small-area estimation, and show that it depends on the intrinsic variability of the target distribution: the value of refinement increases when sub-national distributions differ strongly from the national distribution, and declines as sample size increases.

These findings support a principle: generative models can become powerful inference tools when paired with informative external features. By formalizing and testing this principle across diverse survey settings, our work contributes both a conceptual and a practical framework for using generative models as complements to scarce survey data. It also provides practitioners with empirical guidance on when deploying this framework is likely to add value, particularly in humanitarian contexts where fine-grained evidence is needed but difficult to collect.

\section*{Results}

\subsection*{Context‑conditional generative AI enables sub‑national refinement}
\begin{figure}
\centering
    \includegraphics[width=0.55\textwidth]{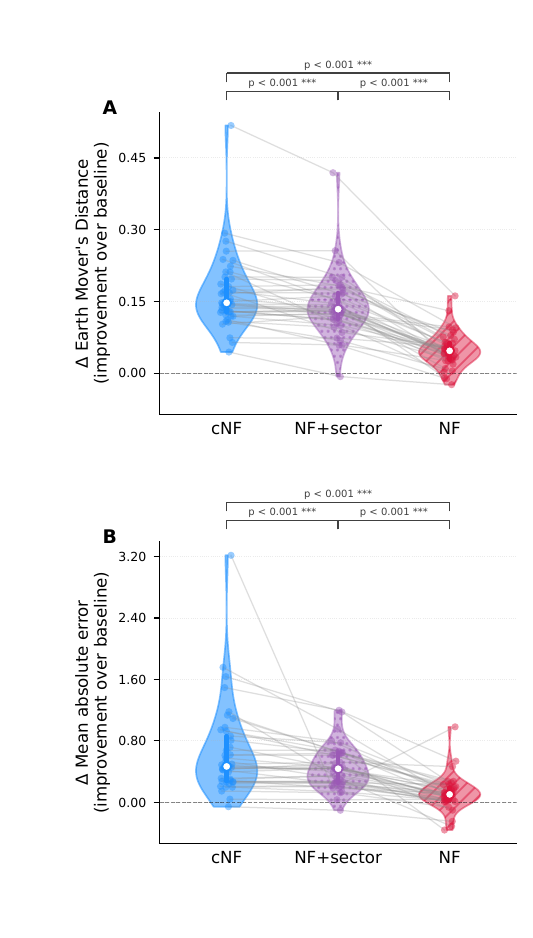}
\caption{\textbf{Context-informed generative models outperform oversampling in generating sub-national data.}
Each point represents a target variable in a dataset, averaged over random seeds and first-level administrative regions. Points are connected across conditioning strategies: full geospatial context (cNF, blue), categorical region and sector (NF+sector, purple), and region only (NF, red). The y-axis shows improvement over oversampling, \(\Delta = Err_{\mathrm{oversampling}} - Err_{\mathrm{model}}\), so positive values indicate better performance. (\textbf{A}) Earth Mover's Distance (EMD) measures distributional similarity on IQR scaled variables; (\textbf{B}) Mean Absolute Error (MAE) measures sub-national mean accuracy in standard-deviation units. Brackets show paired one-sided Wilcoxon signed-rank tests.}
    \label{fig:fig2}
\end{figure}

\begin{figure}
    \centering
    \includegraphics[]{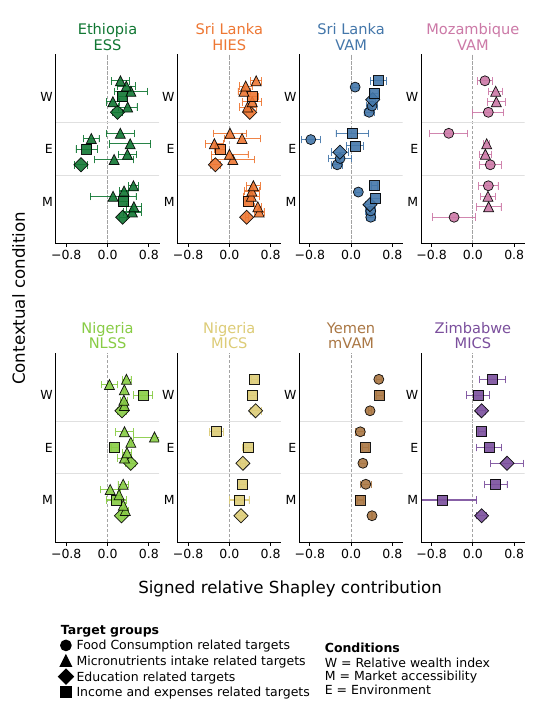}
    \caption{
\textbf{Shapley-based relative contribution of contextual feature groups to synthetic data quality.}
The figure reports signed relative Shapley contributions for each of three contextual feature groups, relative wealth (W), market accessibility (M), and environment (E), across survey datasets and countries. All $2^3 = 8$ subsets are enumerated exhaustively, with a separate cNF model retrained for each subset. Raw Shapley values were normalized by the total absolute Shapley contribution. Points indicate average signed relative Shapley contributions over seeds and first-level administrative units, while horizontal error bars indicate standard errors scaled by the same factor. The vertical dashed line marks zero contribution: values to the right indicate a positive relative contribution to the quality metric, while values to the left indicate a negative relative contribution. Marker shapes distinguish broad groups of outcome variables: food consumption, micronutrient intake, education, and income or expenditure. The specific outcome variables included in each group are reported in the Methods.
}
\label{fig:fig3}
\end{figure}

We compared three conditioning strategies under severe scarcity (one primary sampling unit per first-level administrative unit): full geospatial context (cNF), region plus sector, i.e. settlement type (NF+sector), and region only (NF). Conditioning on a categorical region label (NF) yields only marginal improvement over oversampling, adding sector (NF+sector) yields further gains, and conditioning on the full set of contextual features (cNF) yields systematic performance gains across datasets and generated target variables. Figure~\ref{fig:fig2} quantifies improvement over the oversampling baseline under two complementary metrics: panel~A reports improvement in Earth Mover's Distance (EMD), which measures distributional fidelity at the sub-national level, and panel~B reports improvement in Mean Absolute Error (MAE), which measures the accuracy of sub-national mean estimates. Both panels display the same three conditioning strategies (cNF, NF+sector, NF). Because all three variants share the same flow architecture, the monotonic improvement from NF to NF+sector to cNF indicates that performance gains are driven by the informational content of the conditioning variables.
Unlike Figure~\ref{fig:fig2}, where improvements are measured relative to oversampling, the following Shapley analysis isolates the contribution of continuous contextual information by defining improvement relative to the NF+sector model, thereby removing the effect of sector conditioning and attributing performance gains specifically to additional contextual covariates. We examine the contribution of each contextual feature group to the improvement achieved by the cNF model, quantified using Shapley values. The three groups are relative wealth (W), market accessibility (M), and environment (E). This attribution is quantified via a Shapley value decomposition defined on performance gains rather than on model outputs, distinguishing it from standard SHAP-based interpretability methods. Figure~\ref{fig:fig3} shows that most contextual feature groups contribute positively on average, but contributions vary across datasets and target domains, indicating that no single contextual signal is universally informative. In Ethiopia, wealth- and market-accessibility-related conditions contribute positively and consistently, whereas the contribution of environmental conditions is more mixed. In Nigeria MICS dataset, wealth and market accessibility show a higher contribution, compared to environmental context which shows a less stable contribution. These results suggest that no single covariate dominates globally. Instead, the performance gains of the fully context-informed cNF model reflect the cumulative and dataset-specific signal carried by multiple exogenous contextual features beyond the sector-level conditioning already available to the model. A supplementary analysis suggests that local predictability of target means from contextual features is moderately associated with lower cNF aggregate error (Pearson=$-0.47$, Spearman=$-0.49$; negative correlations indicate that higher local predictability is associated with lower cNF error), but this metric does not fully explain generative gains, which depend on distributional rather than only mean-level structure (Methods, Supplementary Text, Supplementary figure 1, Supplementary figure 2).

\subsection*{Sensitivity to training sample size}
\begin{figure}
    \includegraphics[width=\textwidth]{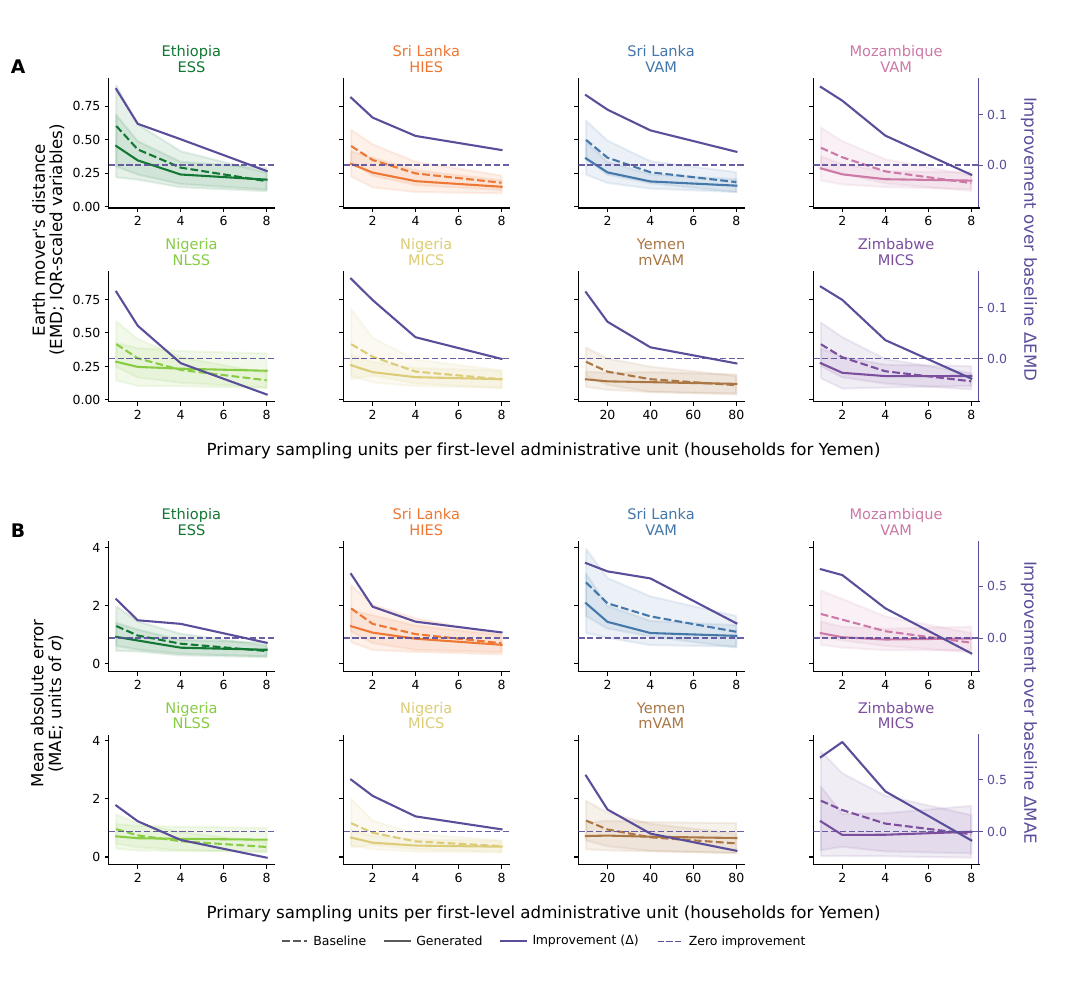}
\caption{
\textbf{Sensitivity analysis of generated data quality to training sample size.}
(\textbf{A}) Earth Mover's Distance (EMD) computed on IQR-scaled variables. (\textbf{B}) Mean Absolute Error (MAE) in units of \(\sigma\). Solid colored lines show the generated model error and dashed colored lines show the oversampling baseline, averaged across variables, first-level administrative units, and random seeds; shaded areas indicate one standard deviation. The secondary y-axis reports improvement over the baseline, defined as \(\Delta = Err_{\mathrm{baseline}} - Err_{\mathrm{generated}}\). The horizontal dashed line marks zero improvement, with positive values indicating better performance than the baseline. Training sample size denotes primary sampling units per first-level administrative unit, except for Yemen, where it denotes households.}
    \label{fig:fig4}
\end{figure}

To assess how training sample size affects the performance gains of the cNF model, we repeated the experiments varying the size of the training subsample. Here, a primary sampling unit (PSU) denotes the survey cluster used as the unit of subsampling; all datasets were subsampled at the PSU level, with the exception of the Yemen mVAM dataset, where households were sampled directly. Consequently, Yemen results should be interpreted as illustrating the same scarcity principle under a different sampling design rather than as directly comparable on the $x$-axis of Figure~\ref{fig:fig4}. Across all datasets, when the subsample is minimal, comprising one PSU per sub-national region, performance improvements are largest and most heterogeneous, reflecting the acute information gap that contextual features help fill (Figure~\ref{fig:fig4}). As the number of PSUs increases, the magnitude of the baseline error itself declines monotonically toward zero, as the subsample becomes increasingly representative of the underlying population. This convergence indicates that the marginal value of contextual features diminishes as direct survey information becomes more abundant.

\subsection*{Improvements are widespread but heterogeneous}
Figure~\ref{fig:fig5} maps the regional difference in EMD between the oversampling baseline and cNF (baseline $-$ cNF) for each dataset and target variable. Across most datasets and targets, improvements are spatially widespread, indicating that cNF generally enhances distributional representativeness relative to the oversampling approach. Variable-level significance, assessed with one-sided paired $t$-tests across regions and seeds with BH-FDR correction, confirms that improvements are significant for nearly all dataset--target combinations, with Ethiopia ESS vitamin B12 as the only non-significant case; this exception is consistent with the very low and nearly spatially invariant vitamin B12 values across the country, which leave little sub-national variation for either the baseline or cNF to capture. At the same time, the maps reveal substantial heterogeneity in these gains, with some regions and variables showing modest gains or occasional reversals where cNF underperforms the baseline. This spatial variation suggests that the benefits of deploying context-conditioned generative AI are not uniform: the same model can substantially improve estimates in some regions while offering limited advantage in others.

\subsection*{Competitive aggregate estimates and improved distributional outputs}
Having shown that cNF improves over oversampling in both distributional and aggregate metrics, we next ask whether these gains remain meaningful when the comparison is made against a stronger baseline designed for small-area mean estimation.{ We compare cNF with a baseline designed for small-area estimation: a Bayesian linear regression fitted on the same contextual features, with a Gaussian prior on the coefficients whose precision, together with the noise variance, is estimated from the data (Methods). Strictly positive outcomes are modelled on the log scale and all others on their original scale. For every surveyed household we draw samples from the Bayesian posterior predictive distribution and generate samples with cNF, and compute all quantities (regional means, local distributions and shares of households at risk) from these samples. The two models therefore differ not in what they output but in what they assume about the conditional distribution: the Bayesian model represents the variation among households that share a context as symmetric noise of nearly constant width, whereas cNF learns its shape, including skewness, bounds and multiple modes, from the data.}

{On regional means the two models perform similarly (Fig.~\ref{fig:fig6}A), with differences distributed across provinces in both directions (Fig.~\ref{fig:fig6}B). However, the models differ markedly in their ability to reproduce the shape. In the example region of Inhambane, Mozambique (Fig.~\ref{fig:fig6}C), the Bayesian model predicts households symmetrically around the regional mean, whereas cNF captures the asymmetry observed in the empirical distributions, such as the concentration of rCSI values near zero. These differences matter for decisions that depend on the tails of a distribution rather than its central tendency. Defining households at risk as those below the national 20th percentile of an outcome (or above the 80th percentile for rCSI and FES), Bayesian misestimates the number of at-risk households more than cNF (Fig.~\ref{fig:fig6}C, right). Across surveys (Fig.~\ref{fig:fig6}D), cNF miscounts fewer at-risk households in six of eight surveys, whereas the Bayesian model performs better in Yemen mVAM and Zimbabwe MICS and for individual variables such as vitamin B12 in Ethiopia ESS.}

{The value of the generative framework therefore lies in distributional detail rather than in aggregate accuracy: cNF performs comparably to the parametric baseline on the aggregate quantities that the baseline is designed to estimate, while additionally providing local distributions from which threshold-based indicators, such as the number of households at risk, can be derived directly. Obtaining comparable distributional estimates from a parametric model requires variable-specific choices of transformation and likelihood functions. In contrast, cNF learns these distributional characteristics directly from the data, without requiring variable-specific model specification. This is particularly useful when small-area descriptions must go beyond aggregate values, for example to identify and characterise households at risk of food insecurity, including in settings where machine learning models are already used to predict outcomes such as the risk of micronutrient inadequacy~\cite{voukelatou2026predicting}.}

\subsection*{Regional heterogeneity favours refinement}
We assess whether the refinement gains achieved by cNF are sufficient to produce sub-national distributions that are more representative than simply assigning the nationally representative distribution from the subsample to each sub-national region. Figure~\ref{fig:fig7} compares three quantities: intrinsic sub-national variability, defined as the EMD between the true (full) national and true (full) sub-national distribution (blue), EMD between true distribution and national subsample used for training (orange), and EMD between true distribution and cNF-generated data (green). Regions are ordered by intrinsic sub-national variability. When intrinsic sub-national variability is low, the national subsample provides a reasonable approximation, and cNF offers no additional benefit. As variability increases, cNF more frequently achieves lower EMD than the national baseline, with the magnitude of improvement growing with variability. This result indicates that generative refinement is most useful in settings where local distributions differ substantially from the national average. However, intrinsic variability as defined here requires knowledge of the true national and sub-national distributions, which is precisely the information unavailable to a practitioner deciding whether to deploy cNF.

To assess whether this circularity can be partially resolved, a supplementary analysis trains a classifier to predict, from \textit{sample-based} proxies alone (intrinsic variability estimated within the sparse subsample, and the regional training sample size) whether cNF refinement yields a positive improvement relative to the national baseline. Evaluated on held-out datasets, the classifier achieves a ROC AUC of 0.67. This suggests limited predictive signal: sample size and intrinsic variability contain some information about when cNF refinement is likely to help, but they do not close the decision loop. Dataset-specific factors that are not reducible to these two quantities still play an important role, and the classifier should not be treated as an operational decision tool.

\subsection*{Distributional improvement does not uniformly translate to regional ranking}
{Humanitarian resource allocation often requires ranking regions by need rather than estimating a single value for each region~\cite{aiken2022targeting}. We therefore test whether the distributional gains reported above translate into a better recovery of the ordering of first-level administrative units. For each dataset, target variable and seed, we compute a regional indicator from the full survey and from each estimate, and measure rank recovery by the Spearman rank correlation between the estimated and true regional indicators (Figure~\ref{fig:fig8}; Methods).}

{The results fall into three groups. In Zimbabwe MICS, Nigeria MICS and Yemen mVAM, cNF improves rank recovery over oversampling, with gains in Spearman correlation between 0.05 and 0.25. These are datasets in which wealth-related targets carry strong spatial structure and the contextual features include a wealth proxy. In Ethiopia ESS, both Sri Lanka surveys and Mozambique VAM, the two methods are indistinguishable. These datasets have 9 to 11 regions, so that the ranking is coarse, and the variability of the correlation across seeds is large. In Nigeria NLSS, cNF recovers the ranking less well than oversampling, with the loss concentrated on micronutrient targets for which the improvement in EMD does not translate into a better recovery of the regional centre.}

{This heterogeneity has a simple explanation. When the error on each regional estimate is comparable in magnitude to the differences between regions, reducing that error does not necessarily recover the ordering. The gains in distributional fidelity are therefore reliable, the gains in regional means are moderate, and the gains in regional ranking depend on the contrast between regions relative to the estimation error.}

These results identify the conditions under which conditional generative refinement is most beneficial for high-resolution sub-national estimation: gains are largest when contextual features are informative, when training samples are small, and when local distributions diverge substantially from the national average, which are the conditions that characterize data-scarce humanitarian settings. Whether these gains carry over to the ordering of regions depends additionally on the contrast between regions relative to the estimation error, which can be partly estimated from the sample before deployment.

\section*{Discussion}

Our results support a simple principle that follows directly from how generative models work: while they cannot create missing information, they can serve as inference tools when informative contextual features enable the recovery of local structure that is poorly represented in the sample. When contextual features encode a signal about sub-national heterogeneity,
a conditional normalizing flow can redistribute that signal across the spatial domain, producing synthetic distributions that are more locally representative than those obtained through mere resampling of the observed data alone. Without informative contextual features, the generative model does not yield consistent improvements and largely reproduces the behavior of simple resampling. This result reflects a generative instantiation of a classical problem in statistical inference: when the covariate distribution at training time differs from the covariate distribution at prediction time, a setting formalised as covariate shift~\cite{shimodaira2000covariate}, standard estimators degrade, and recovery requires leveraging the relationship between features and outcomes to reweight or redistribute information across the target domain. Here, rather than reweighting, the model learns this relationship explicitly and uses it to generate plausible outcomes at poorly represented covariate values. Our framing contributes to the debate on whether generative augmentation of tabular data can ever be of any practical use for inference tasks~\cite{Manousakas2023useful}. The key issue may be less about whether the architecture is sufficiently expressive and more about how to incorporate additional information to support improved inference.

Because all three conditioning variants share the same flow architecture, the monotonic gains from NF to NF+sector to cNF isolate the conditioning signal as the driver: coarse categorical labels allow only group-average distributions, whereas continuous contextual features carry fine-grained spatial information that enables the model to generate plausible outcomes for poorly represented locations. This is the mechanism by which generative models incorporate information: they redistribute structure within the support of the covariate space, rather than extrapolating beyond the support of the outcome space.

\paragraph{Practical implications.}
In practice, the framework is most useful when:
(i)~informative contextual covariates are available,
(ii)~survey samples are sparse at the target spatial resolution, and
(iii)~local distributions differ from national averages.
When these conditions are absent, simpler approaches such as oversampling or direct resampling may be sufficient.

\paragraph{Operational utility.}
The gap between distributional fidelity and regional ranking reported in the Results has direct implications for deployment. A practitioner can expect cNF to improve the local distributional shape reliably, the local mean with moderate confidence, and the local ordering only when the contrast between regions is large relative to the error of each regional estimate. In our data this condition was met for datasets whose targets have strong wealth structure and whose contextual features include a wealth proxy, and it was not met for datasets dominated by micronutrient targets at high regional resolution. This heterogeneity is not a weakness of the approach: it defines the range of settings in which conditional generation adds value for prioritisation, and the quantities that determine it, the intrinsic variability between regions and the generative error, can be partially estimated before any model is used in a decision.

Our objective differs from classical small-area estimation approaches such as Fay--Herriot~\cite{fay1979estimates} or unit-level models~\cite{battese1988error}, including their machine learning extensions~\cite{krennmair2022flexible,corral2025poverty}, which are primarily designed to estimate aggregate indicators such as means, totals, or prevalence rates. In contrast, the proposed framework aims to reconstruct full household-level distributions from which multiple indicators can subsequently be derived. For this reason, we compare aggregate performance against a Bayesian regression baseline while evaluating distributional performance separately through EMD-based metrics: the two comparisons are not competing for the same claim, but establish that cNF matches classical small-area estimators where they are designed to succeed, while additionally recovering distributional information that these approaches do not target. Future work could explore integration with established small-area estimation frameworks.

While context-conditional generative models consistently improve sub-national distributional estimates over simple oversampling, the quality of the training sample remains a prerequisite. Because both cNF and oversampling inherit the support of the training sample, relative improvement over oversampling should not be interpreted as correction of sampling bias. Our framework assumes national-level representativeness with respect to the dimensions relevant to the target variable; generative refinement leverages the information present in the sample but does not correct for its structural deficiencies. A supplementary analysis examines urban-biased sampling, where rural households are systematically excluded (Supplementary Text, Supplementary figure 3). Contextual conditioning can still improve relative to oversampling within the same biased regime, but absolute error for socioeconomic outcomes increases when rural households are excluded.

Several limitations of the current analysis deserve acknowledgment. Our framework assumes that relationships between contextual variables and survey outcomes are sufficiently stable across space. Performance may degrade when these relationships vary systematically across regions. The most consequential manifestation of this assumption is that the covariate--target relationship, $p(Y \mid X)$, is treated as stationary across space. The model learns a single conditional distribution from the pooled national sample and applies it to all regions; if regional ecology, market integration, cultural factors, or conflict exposure give rise to localized relationships that differ from the national average, the generated distributions may be systematically biased in those regions. Evidence from machine learning poverty maps points in the same direction: estimates for unsampled areas are less accurate than for sampled ones, and errors concentrate in poorer and rural areas~\cite{newhouse2023sae,ru2026spatial}. The present evaluation does not directly test this assumption: in the severe scarcity setting every first-level administrative unit contributes at least one PSU to the training sample, so the model always observes some data from each region. A stronger test would be a leave-region-out validation in which one or more administrative units are entirely excluded from training and generation quality is evaluated on those held-out regions. If cNF still improves over the national baseline in this setting, it would provide more direct evidence that the learned conditional transfers across space; if it does not, it would define the spatial scope within which the stationarity assumption is defensible. We consider this an important direction for future work.

A second structural limitation is that the current framework is restricted to continuous target variables because training the cNF model requires continuous outcomes. Several of the targets used here, such as the Food Consumption Score, the reduced Coping Strategies Index, education scores and wealth scores, are discrete or take few distinct values, and the flow does not model this discreteness explicitly. Extending this approach to include mixed-type variables is beyond the scope of this work, but would broaden its practical applicability.

A further limitation concerns the survey design. All surveys use two-stage cluster sampling and provide household-level sampling weights. Our scarcity simulation respects the cluster structure by subsampling at the PSU level and stratifying by region and sector, but the generative model is trained with equal weight on every household and all evaluation metrics are computed on the unweighted sample. The relative comparison between cNF and oversampling is unaffected, since both inherit the same sample, but absolute distances should be read as distances to the survey sample rather than to the population. A weighted evaluation is a natural extension.

Furthermore, the contextual features used for conditioning are themselves model-derived products, often constructed from remote sensing and mobile connectivity data, and may contain substantial measurement error not explicitly accounted for in our framework. The Shapley analysis confirms that contextual variables occasionally show negative contributions for specific dataset--target combinations, indicating that external context can introduce noise rather than useful signal. Future work could extend the conditioning set with additional humanitarian covariates such as conflict exposure, displacement patterns, and market disruptions. A related limitation is that household geolocation data are typically coarsened to protect privacy, so conditioning information remains spatially aggregated and cannot fully capture micro-level heterogeneity. Conversely, generating household-level records conditional on fine-grained location features could raise re-identification risk; any operational deployment should assess disclosure risk relative to the spatial resolution and population density of the target domain.

A further limitation arises from the evaluation of model performance. Although the model is trained on the joint distribution of all variables, the metrics are computed separately for each variable. As a result, errors in the dependence between variables are not captured, even though such relationships may be important for joint outcomes.

Finally, the empirical scope of this study is limited to humanitarian household surveys. Although the datasets span multiple countries and outcome domains, they share broadly similar survey structures. The same logic may be relevant to other settings where outcomes are spatially structured and direct measurement is sparse, including epidemiological surveillance, poverty mapping, and agricultural yield estimation, but generalization will require dedicated empirical validation. Looking forward, a promising development is the emergence of foundation models that encode multi-modal descriptions of place into task-agnostic spatial embeddings that can serve as conditioning inputs for downstream models~\cite{agarwal2024pdfm}.

Our results show that context-conditioned generative models can extract additional inferential signal from sparse survey data when external contextual features are informative about local variation. Household surveys are a case in point: direct measurement is costly, spatial coverage is often uneven, and decisions frequently require estimates below the level at which samples are fully representative. At the same time, our findings also clarify the limits of this approach. Generative refinement is most useful when the available sample is insufficiently representative at the target spatial resolution, when local distributions differ meaningfully from the national distribution, and when contextual features capture part of that local variation. When these conditions are not met, simpler baselines may perform similarly. The framework developed here, therefore, provides a practical basis for assessing when contextual information can yield reliable inferential gains and when generative augmentation is unlikely to justify its added complexity.

\section*{Methods}

\subsection*{Datasets}
\label{sec:datasets}

We evaluate our approach on eight household survey datasets spanning diverse geographic and socioeconomic contexts (Supplementary Table 1).
All surveys use a two-stage stratified cluster sampling design in which households are nested within primary sampling units (PSUs), the operational units deployed in the field.
Each dataset provides household-level observations of target variables, together with geographic identifiers at the first- or second-level administrative resolution.
\paragraph{Ethiopia Socioeconomic Survey 2018--2019 (ESS).}
We use the fourth wave of the Ethiopia Socioeconomic Survey \cite{ess_eth_2018_2019}, a nationally representative survey that collected information from 6{,}770 households between May and September 2019.
We extract seven household-level variables: the average education level of adults, the logarithm of non-food expenditures, and estimates of apparent micronutrient intake for five micronutrients (iron, Folate ($\mu$g), Vitamin A ($\mu$g RAE), Vitamin $B_{12}$ ($\mu$g), Zinc (mg)), extracted maintaining consistency with methods established in prior research \cite{tang2022modeling, voukelatou2026predicting}.
After preprocessing, the analytical sample comprises 6{,}214 households distributed across 11 first-level administrative units.

\paragraph{Nigerian Living Standards Survey 2018--2019 (NLSS).}
We use the Nigerian Living Standards Survey \cite{nlss_nga_2018_2019}, a nationally representative survey that collected information from 22{,}587 households between September 2018 and September 2019.
We extract the same seven variables as for the ESS \cite{tang2022modeling, voukelatou2026predicting}.
The analytical sample comprises 22{,}106 households distributed across 37 first-level administrative resolution units.

\paragraph{Sri Lanka Household Income and Expenditure Survey 2019 (HIES).}
We use the Sri Lanka HIES 2019 \cite{hies_lka_2019}, which was conducted over 12 months to capture seasonal variation, with a total sample size of 25{,}000 housing units.
Sri Lanka comprises 9 first-level administrative divisions and 25 second-level administrative divisions.
We extract seven household-level variables: the average education level of members aged 17 or older, the logarithm of household income, and apparent intake for five key micronutrients, as for the ESS and NLSS \cite{tang2022modeling, voukelatou2026predicting}.

\paragraph{Sri Lanka Vulnerability Analysis and Mapping survey 2024 (VAM).}
We use the WFP Vulnerability Analysis and Mapping survey for Sri Lanka \cite{vam_lka_2024}, which randomly selects 600 households per district for a total of 15{,}000 households across 25 districts.
We extract six variables: an education score (average of the fraction of adults with higher education and the fraction of children attending school), the logarithm of household income, the reduced Coping Strategies Index (rCSI), the Food Consumption Score (FCS), the Food Expenditure Share (FES), and space per person (number of rooms divided by household size).

\paragraph{Yemen mobile Vulnerability Assessment and Mapping 2025 (mVAM).}
We use the Yemen mVAM dataset collected between August 2025 and January 2026 \cite{wfp_mvam_yemen_2025_dataviz_assessments}, gathered by WFP through daily Computer Assisted Telephone Interviews.
The analytical dataset contains 54{,}702 observations from 44{,}675 distinct households, distributed across 22 first-level administrative resolution units and 331 second-level administrative resolution units.
We extract three variables: the logarithm of total monthly per-person expenditures, the rCSI, and the FCS.

\paragraph{Mozambique Post-Shock Food Security Assessment 2023.}
We use the Mozambique Post-Shock Food Security Assessment \cite{wfp_acr_mozambique_mz02_2023}, which covers 11{,}080 households across 11 first-level administrative resolution units and 72 second-level administrative resolution units.
We extract four variables: the FCS, the rCSI, and two composite scores representing the weekly frequency of consumption of vitamin~A-rich and protein-rich each based on household-reported days of consumption in the preceding week (range 0--7).

\paragraph{Nigeria Multiple Indicator Cluster Survey (MICS).}
We use the Nigeria Multiple Indicator Cluster Survey 2021, Round 6 \cite{nga_mics_2021}, conducted jointly by the National Bureau of Statistics and UNICEF. The survey sampled 41,532 households across 37 states, of which 39,632 households were successfully interviewed, together with 38,806 women and 17,347 men aged 15--49. Nigeria comprises 37 first-level administrative units.
We extract three household-level variables: the average adult education level, space per person, and a composite wealth score.

\paragraph{Zimbabwe Multiple Indicator Cluster Survey (MICS).}
We use the Zimbabwe Multiple Indicator Cluster Survey 2019, Round 6 \cite{zwe_mics_2019}, conducted by the Zimbabwe National Statistics Agency with support from UNICEF. A total of 12,012 households were selected across 462 clusters; 11,091 households were successfully interviewed. The survey was designed to provide estimates at the national level and for ten provinces: Bulawayo, Manicaland, Mashonaland Central, Mashonaland East, Mashonaland West, Matabeleland North, Matabeleland South, Midlands, Masvingo, and Harare.
We extract three household-level variables: the average adult education level, space per person, and a composite wealth score.

\subsection*{Problem formulation}

We study the problem of estimating sub-national distributions and aggregate statistics from sparse household survey data. Let $Y \in \mathbb{R}^d$ denote a vector of continuous target variables of interest (e.g., food security or socioeconomic indicators). Let $\ell \in \{0, 1, \ldots, L\}$ denote a level of administrative resolution, where $\ell = 0$ corresponds to national level, $\ell = 1$ to first-level administrative resolution and $\ell = L$ to the finest available level. At each level, let $a_\ell \in \mathcal{A}_\ell$ denote the sub-national region index of a given household\footnote{For privacy protection, household locations are typically available only at an aggregated spatial resolution, usually up to the first- or second-level administrative unit}, where $\mathcal{A}_\ell$ is the set of regions at level $\ell$. Let $s \in S$ denote the sector a given household belongs to (rural, urban, estate)\footnote{Estate sector is present only in Sri Lanka}, and let $X \in \mathbb{R}^p$ denote a vector of contextual features (e.g., accessibility, wealth proxies, environmental indicators).

Our inferential target is the regional distribution $p(Y \mid a_1)$: the distribution of household-level outcomes within each first-level administrative unit, aggregated over sector. During model training, sector membership and contextual covariates are incorporated as conditioning information, so that the generative model learns distributions of the form $p(Y \mid a, s, X)$. Regional distributions are then obtained by aggregating over the known sector composition:
\begin{equation}
    p(Y \mid a) = \sum_{s} p(Y \mid a, s, X_{a})\, p(s \mid a),
\end{equation}
where $p(s \mid a)$ is treated as known from the survey design. Sector is therefore not part of the target but part of the conditioning structure used to improve estimation; its regional composition is typically available from survey metadata or census data. All evaluations are performed at the first-level administrative resolution by comparing generated and observed regional distributions $p(Y \mid a_1)$.

Consider a limited sample of household observations $\{(Y_i, a_{L,i}, s_i)\}_{i=1}^n$ that is representative only at national level $\ell=0$ with respect to spatial and sector dimensions, and can therefore only be used to estimate $p(Y \mid a_0)$. By enriching this sample with contextual features, so that it becomes $\{(Y_i, X_{a_{L,i}}, s_i)\}_{i=1}^n$, and assuming that the relationship between $X$ and $Y$ is homogeneous across space, we train a generative model to learn $p(Y \mid X, s)$. This learned conditional can then generate synthetic household-level data for any region $a_{\ell}$ for which contextual features are available, by sampling $Y_i \sim p(Y \mid X_{a_{\ell}}, s)$.

\subsection*{Data scarcity simulation}
\label{sec:scarcity}

We treat each full survey dataset $D=\{(Y_i, a_{L,i}, s_i)\}_{i=1}^m$, representative at the sub-national level by design through two-stage cluster sampling~\cite{WFPsampling}, as ground truth. From $D$ we construct a sparse training sample $D'$ by subsampling at the primary sampling unit (PSU) level, since PSUs are the operational units through which household surveys are implemented in the field. PSU selection is stratified within each first-level administrative unit so that the rural--urban composition of $D'$ approximates that of $D$; the resulting subsample therefore preserves national-level spatial and sectoral composition while, by construction, destroying sub-national representativeness. In the most severe scarcity setting, one PSU is selected per first-level administrative unit (typically ${\sim}10$ households per PSU, with variation across clusters). The Yemen mobile VAM dataset is an exception: since mobile surveys do not rely on PSUs, households are sampled directly.

We denote by $\tilde{D}=\{(Y_i, X_{a_{L,i}}, s_i)\}_{i=1}^{m-n}$ the held-out observations excluded from $D'$. Contextual features are attached to all households at the finest available administrative resolution, so that $D'$ becomes $\{(Y_i, X_{a_{L,i}}, s_i)\}_{i=1}^n$. A conditional normalizing flow is trained on the enriched $D'$ and used to generate target values for the held-out locations: $Y_{j}^{\mathrm{gen}}\sim p(Y|X_{a_{\ell}},s)$. The generated set $G=\{(Y_i^{\mathrm{gen}}, X_{a_{L,i}}, s_i)\}_{i=1}^{m-n}$ is evaluated against $\tilde{D}$ by comparing distributions and aggregate means at the first-level administrative resolution. All subsampling steps are repeated with five independent random seeds; results in figures~\ref{fig:fig2}--\ref{fig:fig8} are averaged across seeds. Figure~\ref{fig:fig1} illustrates the framework.

\subsection*{Sample size sensitivity}
\label{sec:sensitivity}

To characterise robustness under varying degrees of data scarcity, we repeat all experiments while varying the number of PSUs selected per first-level administrative resolution region from 1 (extreme scarcity) to 8 (moderate scarcity).
All methods are evaluated under identical sampling conditions for each setting.
For each level of data scarcity (PSU count), we compute performance improvements across many scenarios (regions, variables, seeds), and then summarise them into a single average trend. To capture variability in performance, we also report one standard deviation across regions and variables. These results are summarised in Figure~\ref{fig:fig4}.

\subsection*{Contextual features}
\label{sec:features}

We incorporate continuous contextual features that capture environmental and socioeconomic conditions expected to explain sub-national variation in the target variables. All features are matched to household locations using the finest available spatial resolution and are treated as exogenous predictors throughout. We also incorporate a categorical feature describing the settlement sector of a given household.

\paragraph{Market accessibility.}
We use estimated travel time to the nearest market or urban centre, computed at high spatial resolution from road-network and land-cover data~\cite{benassai2025unequal}.
This variable captures remoteness as a proxy for access to food markets and services.

\paragraph{Relative Wealth Index (RWI) and wealth indicator (wscore).}
We use the RWI derived from de-identified mobile-phone connectivity data and satellite imagery~\cite{chi2022microestimates}.
The index provides a continuous measure of relative household wealth at sub-kilometre resolution. For Yemen, the relative wealth index dataset was not available, so we substituted it with a wealth indicator \textit{wscore} coming from the most recent MICS survey, available at the first administrative level.

\paragraph{Environmental indicators.}
We incorporate the Normalised Difference Vegetation Index (NDVI), land surface temperature, and precipitation summaries aggregated over relevant seasonal windows~\cite{huang2021commentary}.
These variables capture agro-climatic conditions that influence food production and dietary outcomes, and are included primarily for datasets containing micronutrient intake targets.

\paragraph{Sector.}
The sector variable identifies the settlement or survey-sector category associated with each observation. In all datasets, sector distinguishes rural from urban areas; in the Sri Lankan surveys, an additional estate sector is also included, reflecting the country-specific sampling structure. This variable captures broad differences in population setting that are part of the survey design and may be associated with differences in livelihoods, infrastructure, market access, and household characteristics.

\subsection*{Conditional normalizing flow model}
\label{sec:model}

We model $p(Y \mid X)$ using conditional normalizing flows (cNF).
A normalizing flow defines an invertible, differentiable mapping $f_\theta : \mathbb{R}^d \to \mathbb{R}^d$ from the target space to a latent space with a tractable base distribution $p_Z$ (here, a standard multivariate normal).
Conditioning on $X$ is incorporated by making the parameters of each coupling layer a function of $X$:
\begin{equation}
    Z = f_\theta(Y;\, X),
\end{equation}

with conditional likelihood
\begin{equation}
    \log p(Y \mid X) = \log p_Z\!\left(f_\theta(Y; X)\right)
                     + \log \left|\det \frac{\partial f_\theta}{\partial Y}\right|.
\end{equation}
Model parameters $\theta$ are estimated by maximising the conditional log-likelihood over the observed training data.
At inference time, synthetic samples for region $a_{\ell}$ are generated by drawing $Z \sim p_Z$ and applying the inverse transformation conditioned on the contextual features $X_a$ associated with that region:
\begin{equation}
    Y = f_\theta^{-1}(Z;\, X_{a_{\ell}}).
\end{equation}
We use the publicly available implementation in the Python library \texttt{probaforms} \cite{hushchyn2023probaforms}.

\paragraph{Feature scaling.}
Target variables are standardised before training by subtracting the median and dividing by the interquartile range of the training subsample, and generated values are mapped back to the original units with the inverse transformation. Continuous contextual features are passed to the model in their original units, concatenated with the one-hot sector encoding. Standardising the conditioning vector is common practice for neural density estimators and would be expected to improve rather than degrade the reported gains, so the results presented here should be read as a conservative estimate of cNF performance.

\paragraph{Context-conditioned variant (cNF).}
In the primary setting, the model is conditioned on the full set of contextual features available for a given dataset, as described in the previous section. The sector condition is one-hot encoded and attached to the continuous part of the conditional input.
These features provide the information used to refine sub-national distributions beyond what is available in the sparse observed sample.

\paragraph{Region-conditioned variant (NF).}
To isolate the contribution of continuous context variables, we consider a reduced model conditioned only on a one-hot encoding of the first-level administrative resolution region label.
This variant retains spatial grouping but removes all exogenous contextual features, allowing a direct assessment of how much of the improvement is attributable to informative conditioning versus the sole region of belonging identification.

\subsection*{Baselines}

In different parts of this work, we compare the cNF and NF models to three reference methods.

\paragraph{Oversampling.}
For each held-out record, the oversampling baseline draws a target vector with replacement from the pooled national sparse subsample, ignoring all contextual covariates and regional labels. In the main comparison to cNF, oversampling is evaluated at the first-level-administrative resolution after matching the held-out sample size. Because the resampling pool is the entire national subsample and no conditioning information is used, every region receives draws from the same national distribution. This baseline serves as the primary reference for evaluating whether the generative framework leveraging exogenous information can outperform structured resampling.

\paragraph{National sample baseline.}
Regional distributions are approximated by assigning the same national-level distribution from the subsample to all regions, thereby ignoring any sub-national heterogeneity. This baseline serves to assess whether the conditional generative model provides meaningful improvements for small-area estimation.

\paragraph{Bayesian multivariate linear baseline.}
{In the context of small-area estimates, we use this baseline to assess the performance of our method for aggregate statistics, specifically average estimates, and for local distributions.
We implement a multivariate Bayesian linear regression with a Gaussian prior on the regression coefficients, which is the standard "Bayesian ridge regression" formulation typically used in practice for this class of models.
Let $Y_i \in \mathbb{R}^q$ denote the vector of target variables for observation $i$,
and $X_i \in \mathbb{R}^p$ the corresponding vector of conditioning features (plus an intercept term).
For each target variable $j = 1, \dots, q$, the model is specified independently as:
\[
Y_{ij} = X_i^\top \beta_j + \epsilon_{ij}, \qquad \epsilon_{ij} \sim \mathcal{N}(0, \sigma_j^2),
\]
with a zero-mean Gaussian prior on the coefficients, $\beta_j \sim \mathcal{N}(0, \lambda_j^{-1} I)$, and Gaussian noise precision $\alpha_j = \sigma_j^{-2}$. Targets that are strictly positive and not already expressed in logarithms are modelled on the log scale, i.e.\ $Y_{ij}$ is replaced by $\log Y_{ij}$. Both the prior precision $\lambda_j$ and the noise precision $\alpha_j$ are estimated from the data by evidence maximization, rather than fixed a priori, yielding a closed-form posterior.
For a new input $X_i$, the posterior predictive distribution for target variable $j$ is Gaussian,
\[
\mathcal{N}\bigl(\hat{\mu}_{ij}, \hat{\sigma}_{ij}^2\bigr),
\]
where $\hat{\mu}_{ij} = X_i^\top \hat{\beta}_j$ and $\hat{\sigma}_{ij}^2$ is the posterior predictive variance, which jointly accounts for the estimated noise variance and the posterior uncertainty over $\beta_j$.
For each conditioning vector, the Bayesian model thus produces a Gaussian posterior predictive distribution for every target variable (log-normal on the original scale for log-transformed targets), characterised by a mean $\hat{\mu}_{ij}$ and standard deviation $\hat{\sigma}_{ij}$. Because the model assumes Gaussian outcomes, these two quantities fully characterise the predictive distribution. For each surveyed household we draw samples from its posterior predictive distribution, and regional means, distributions and shares of households at risk are computed from these samples within each first-level administrative unit, using the same procedure as for the cNF-generated data. We compare the resulting regional estimation to those obtained from the cNF-generated data and from the ground-truth survey. We also tested a richer specification in which the continuous contextual features were expanded with quadratic terms and pairwise interactions, with the same prior. Under the data scarcity setting considered here, this specification led to overfitting and performed worse than the linear model, so the linear specification is used throughout.}

\subsection*{Evaluation metrics}
\label{sec:metrics}

\paragraph*{Earth Mover's Distance (EMD).}
Distance between distributions is measured using the Earth Mover's Distance (Wasserstein-1 distance) between the generated and true sub-national distributions.
For two univariate distributions $P$ and $Q$, $\text{EMD}(P,Q) = \int |F_P(x) - F_Q(x)|\,dx$, where $F_P$ and $F_Q$ are the respective CDFs.
Lower EMD indicates higher distributional fidelity. Target variables are scaled by the interquartile range (IQR) of their full-dataset distribution prior to distribution evaluation, so that metrics are comparable across variables with different units and magnitudes. IQR scaling is preferred over standard-deviation scaling because several target variables, particularly expenditure and consumption measures, exhibit heavy tails for which the standard deviation is sensitive to extreme values. Note that EMD is therefore reported in IQR units, whereas MAE (below) is expressed in standard-deviation units; the two metrics are not on a common scale.

\paragraph*{Mean Absolute Error (MAE)}
Aggregate accuracy is measured as the absolute difference between the estimated and true sub-national mean, expressed in units of the target variable's standard deviation:
\begin{equation}
    \text{MAE}_r = \frac{|\hat\mu_r - \mu_r|}{\sigma_Y}.
\end{equation}

\paragraph*{Improvement over baseline.}
To quantify improvement relative to oversampling, we define
\begin{equation}
    \Delta = \text{Err}_{\text{baseline}} - \text{Err}_{\text{model}},
\end{equation}
where Err is either EMD or MAE.
Positive $\Delta$ indicates that the model outperforms oversampling; $\Delta = 0$ corresponds to parity.
Results are always averaged on random seeds.

\subsection*{Rank recovery}
\label{sec:rank}

To assess whether improvements in distributional fidelity translate into a better ordering of regions, we compute for each dataset, target variable and seed a regional indicator from the full survey and from each estimate. The indicator is the share of households in the region whose value falls below the national first quartile of the target variable, computed on the full survey for that seed; results using the regional median in place of this share are similar. The indicator is computed from the sparse subsample directly, from the oversampling baseline, and from the cNF-generated data, in each case restricted to the regions for which all three estimates are available. Rank recovery is measured by the Spearman rank correlation between the estimated and true regional indicators. It is averaged over seeds and reported with their its standard error.

\subsection*{Statistical tests}
\label{sec:tests}

\begin{figure}
    \includegraphics[width=0.9\textwidth]{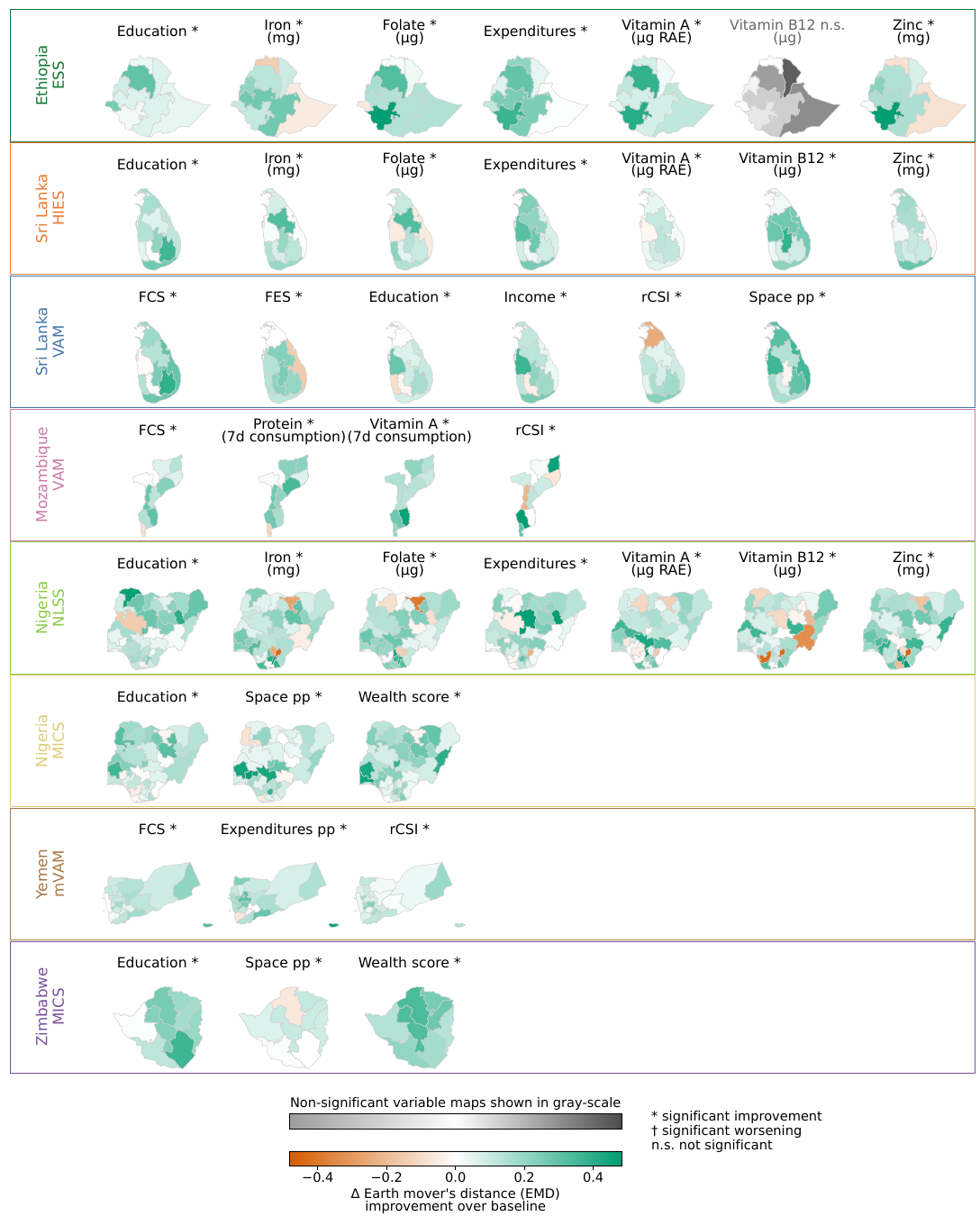}
\caption{\textbf{Spatial heterogeneity in distribution refinement.}
Maps show $\Delta$EMD between the oversampling baseline and the cNF model at first-level administrative resolution, on IQR-scaled variables and averaged over seeds. Positive values indicate better cNF performance; negative values indicate better oversampling performance. Coloured maps indicate variables with significant improvement, tested with a one-sided paired $t$-test across regions and BH-FDR correction; non-significant variables are shown in greyscale. Symbols denote significant improvement (*), significant worsening (\dag), and non-significance (n.s.).}
    \label{fig:fig5}
\end{figure}

\begin{figure}
\centering
    \includegraphics[width=0.85\textwidth]{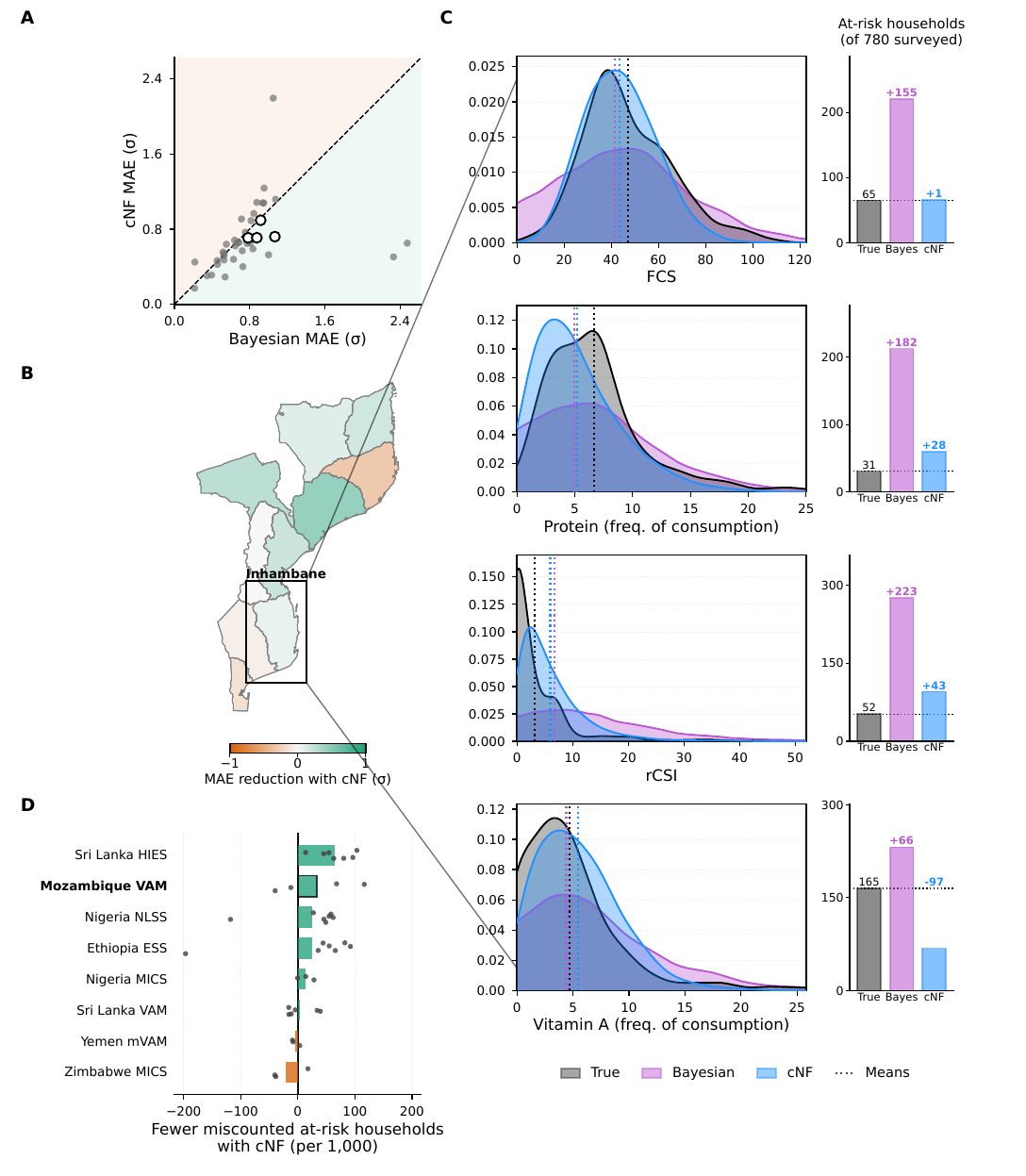}\\
\caption{
\textbf{Aggregate error and distributional comparison, with local diagnostics.}
\textbf{(A)} Mean absolute error (MAE) of regional means for the Bayesian baseline ($x$ axis) and the context-conditioned normalizing flow (cNF, $y$ axis), in target-variable standard deviation units. Each point is one dataset--target-variable combination; points below the diagonal favour cNF; open circles are the Mozambique VAM variables shown in \textbf{B} and \textbf{C}.
\textbf{(B)} Mozambique VAM map, coloured by mean MAE improvement of cNF over the Bayesian baseline, $\Delta = \mathrm{MAE}_{\mathrm{Bayesian}} - \mathrm{MAE}_{\mathrm{cNF}}$. Box: Inhambane.
\textbf{(C)} Household-level distributions for Inhambane: true, Bayesian posterior predictive, and cNF-generated, with dotted lines at each mean; bars give the at-risk households among those surveyed (below the national 20th percentile, above the 80th for rCSI) and each model's over- or under-count.
\textbf{(D)} Difference in miscounted at-risk households (per 1,000 households) between the Bayesian baseline and cNF, per survey (bars) and variable (dots); positive values favour cNF.
}
    \label{fig:fig6}
\end{figure}

\begin{figure}
    \begin{minipage}{\linewidth}
        \includegraphics[width=\textwidth]{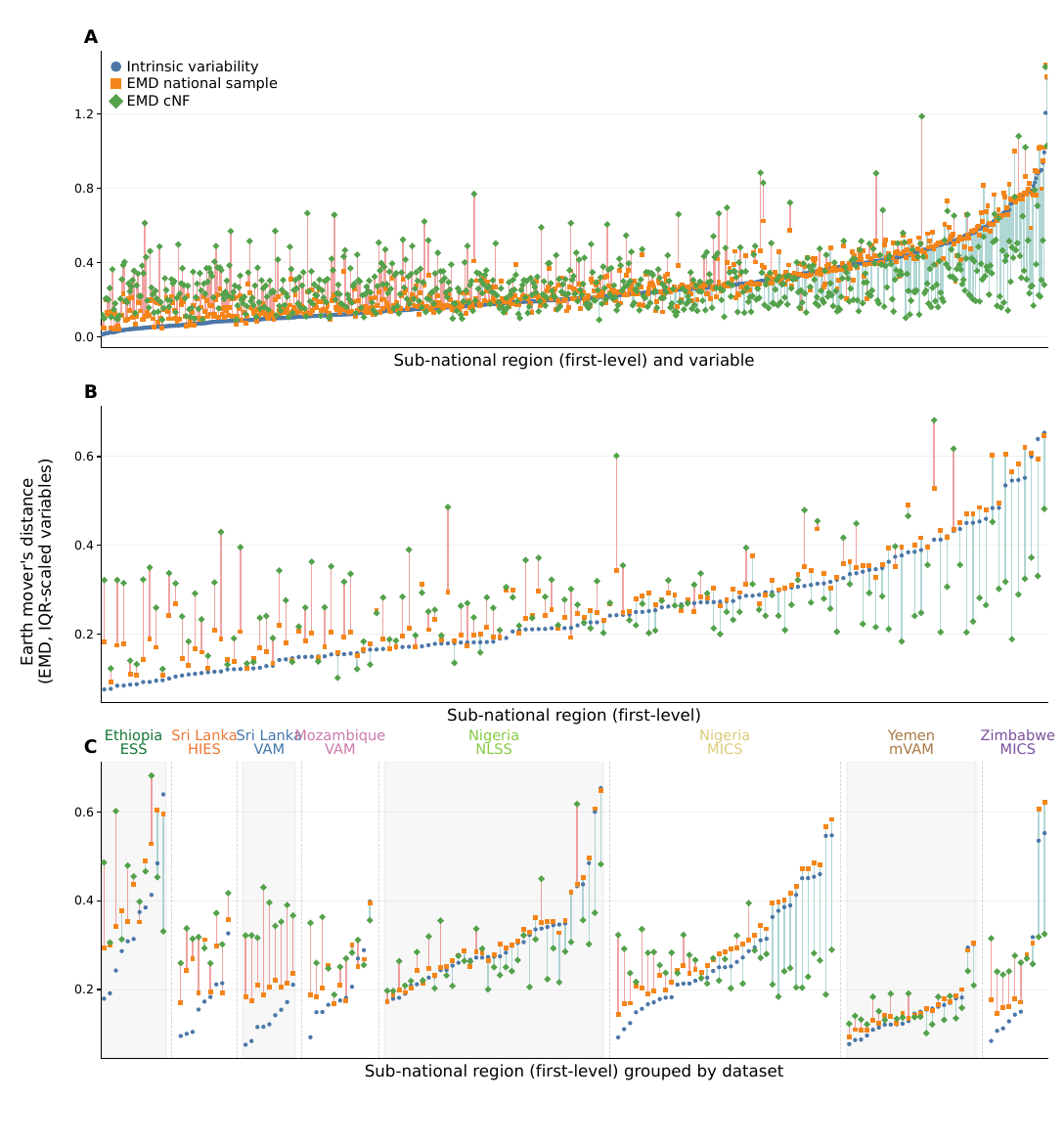}
    \end{minipage}
\caption{\textbf{Context-conditioned normalizing flow improves sub-national estimates when regional variability is high.}
Points show Earth Mover’s Distance (EMD), averaged over seeds. In panel A, each point corresponds to one variable in one first-level administrative unit; in panels B and C, values are also averaged over variables. Blue circles indicate intrinsic variability in the true data, orange squares the national subsample baseline, and green diamonds the cNF-generated data. Vertical connectors are blue when cNF improves over the baseline and red otherwise. \textbf{A}: regions and variables ordered by intrinsic variability. \textbf{B}: regions averaged over variables and ordered by intrinsic variability. \textbf{C}: same regions grouped by dataset.}

    \label{fig:fig7}
\end{figure}

\begin{figure}
\centering
    \includegraphics[width=0.75\textwidth]{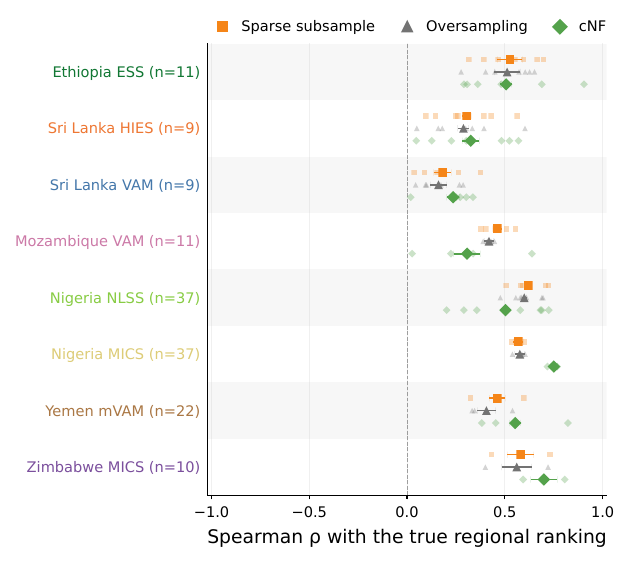}
\caption{\textbf{Rank recovery over the regions.}
Spearman rank correlation between the true regional (first level administrative unit) indicator, computed on the full survey, and the indicator estimated: from the sparse subsample used directly (orange squares), from the oversampling baseline (gray triangles), and from the cNF-generated data (green diamonds), for each dataset under severe scarcity. Large markers show the mean over seeds with its standard error; small markers show individual seeds and target variables. The number of first-level administrative units in each dataset is given in parentheses.}
    \label{fig:fig8}
\end{figure}

We use paired one-sided Wilcoxon signed-rank tests to compare model performance across dataset--target-variable pairs, where each pair contributes one observation averaged over regions and seeds. This test is used in Figure~\ref{fig:fig2} (cNF vs.\ NF+sector vs.\ NF vs.\ oversampling) and Figure~\ref{fig:fig6}A (cNF vs.\ Bayesian baseline). For the spatial analysis in Figure~\ref{fig:fig5}, variable-level significance is assessed with one-sided paired $t$-tests across first-level administrative regions (averaged over seeds), with Benjamini--Hochberg false discovery rate (BH-FDR) correction applied across variables within each dataset. In the supplementary bias analysis (Supplementary figure 3), group differences are assessed with two-sided Welch's independent two-sample $t$-tests. Test in Figure~\ref{fig:fig5} uses $\alpha=0.10$, all remaining tests use $\alpha = 0.05$. Rank recovery (Figure~\ref{fig:fig8}) is reported as means and standard errors over seeds without a formal test, since the number of regions per dataset is small. Because multiple target variables within a dataset are not fully independent, the resulting $p$-values should be interpreted as descriptive measures of evidence rather than exact estimates under strict independence assumptions; the consistency of effects across datasets, targets, and evaluation metrics provides additional support for the robustness of the observed patterns.

\subsection*{Intrinsic sub-national variability}
\label{sec:variability}

We measure intrinsic sub-national variability as the EMD between the true national distribution and the true first-level administrative unit distribution.
Regions with high intrinsic variability deviate substantially from the national average; those with low variability are well-approximated by the national subsample.
This measure is used to assess the utility of cNF for small-area refinement in Figure~\ref{fig:fig7}.

\subsection*{Feature attribution via Shapley values}
\label{sec:shapley}

To assess the contribution of individual contextual variables to the improvement achieved by the fully context-informed cNF model, we compute Shapley values~\cite{lundberg2017unified} for the performance gain \(\Delta\). In this analysis, \(\Delta\) measures the added value of exogenous geospatial context relative to the NF+sector model, rather than relative to the oversampling baseline. We define \(\Delta = Err_{\mathrm{NF+sector}} - Err_{\mathrm{cNF}}\), so that positive values indicate that adding the full set of contextual variables improves synthetic data quality beyond conditioning on sector alone. The Shapley baseline therefore includes sector conditioning: the decomposition isolates the added contribution of the exogenous continuous features, not their total effect.
For each combination of dataset, target variable, first-level administrative region, and random seed, the Shapley value of contextual feature group \(X_j\) is estimated by averaging the marginal change in \(\Delta\) when \(X_j\) is included versus excluded, over all possible subsets of the remaining groups. Because only three contextual feature groups were considered, exact Shapley values were computed by evaluating all possible feature-group subsets; for each subset, a separate cNF model is retrained using only that subset of contextual feature groups (together with sector and region), so that Shapley contributions reflect the effect of each group on the learned conditional distribution rather than on inference-time masking or permutation. Shapley values are computed at the level of individual regions, target variables, and seeds, and are then averaged; uncertainty is reported as the standard error of the mean across regions and seeds. Positive Shapley values indicate contextual feature groups that increase the advantage of the fully context-informed cNF model over NF+sector, while negative values indicate groups whose inclusion reduces this advantage. Negative contributions can arise from collinearity among contextual feature groups, from measurement error in model-derived covariates, or from cases where a group introduces noise for a particular dataset--target combination rather than useful signal. For visualization, target variables are grouped into four broad domains. Food-consumption-related targets include the Food Consumption Score (FCS), Food Expenditure Share (FES), and reduced Coping Strategies Index (rCSI). Micronutrient-intake-related targets include iron (mg), folate (\(\mu\)g), vitamin A (\(\mu\)g retinol activity equivalents, RAE), vitamin B12 (\(\mu\)g), and zinc (mg), together with food-group indicators related to weekly consumption of protein and vitamin A. Education-related targets include adult education and education score. Income- and expenditure-related targets include log expenditures, log per-capita expenditures, log income, wealth score, and space per person.

\subsection*{Decision support meta-model}

To characterize when generative refinement is most beneficial, we trained a supplementary decision-tree model using region-level features as predictors.
For each region, the input features were the intrinsic variability of the target distribution, measured as the mean discrepancy between the regional ($\ell=1$) and national distributions within the training subsample, and the effective training sample size.
We defined a binary target equal to 1 when the mean improvement, defined as the difference in EMD between the baseline and the cNF, was positive and 0 otherwise, and trained a \texttt{DecisionTreeClassifier} with balanced class weights. Performance was evaluated with leave-one-dataset-out cross-validation, and hyperparameters were tuned via stratified 5-fold cross-validation within each training set. Classification performance was evaluated using ROC AUC on the held-out datasets.
This setup evaluates whether a simple interpretable model can predict, based solely on sample size and intrinsic variability, when cNF refinement is likely to yield more accurate distributions than the national subsample baseline on datasets that are scarce to begin with.

\newpage

%%%%%%%%%%%%%%%% MAIN TEXT FIGURES %%%%%%%%%%%%%%%

% flowchart figure

% violin plots

%shapley

% sensitivity and context

% maps

% bayesian comparison with local diagnostics

% dumbell national intrinsic

% rank recovery

\clearpage

\bibliography{biblio}
\bibliographystyle{naturemag}

%%%%%%%%%%%%%%%% ACKNOWLEDGEMENTS %%%%%%%%%%%%%%%

\section*{Acknowledgments}
We would like to thank Frances Knight for the support and feedback, Jonas De Meyer for the support on extracting and interpreting the mVAM data, and Shehan Fernando for the support on interpreting the Household Income and Expenditure Survey 2019 (HIES 2019).
We acknowledge the Department of Census and Statistics, Sri Lanka, for providing the Sri Lanka HIES 2019 dataset used in this analysis.

\paragraph*{Funding:}
F.S., K.Kalimeri, D.Paolotti, R.S. acknowledge support from the Lagrange Project of the ISI Foundation, funded by Fondazione CRT.
V.V., D.Piovani, and K.Koupparis were supported by the Gates Foundation (INV-037325) through the Modelling \& Mapping the Risk of Inadequate Micronutrient Intake (MIMI) project.
K.K. additionally received support from the Spanish Agency for International Development Cooperation (AECID) via UNICEF’s Frontier Data Network.

\paragraph*{Author contributions:}F.S. contributed to conceptualization, methodology, software, formal analysis, visualization, and writing of the original draft. V.V. contributed to conceptualization, formal analysis, data curation, and writing of the original draft. D. Piovani, K. Koupparis, D. Paolotti and R.S. contributed to conceptualization and formal analysis. K. Kalimeri contributed to conceptualization, formal analysis, and writing of the original draft. All the authors have read and approved the final manuscript.

\paragraph*{Competing interests:}
There are no competing interests to declare.

\paragraph*{Code availability:}

The code used to preprocess the data, train the models, reproduce the analyses, and generate the figures is available at \cite{sibilla2025code}.

\paragraph*{Data availability:}
Publicly available survey datasets used in this study, including the Ethiopia ESS, Nigeria NLSS, Nigeria MICS, and Zimbabwe MICS datasets, can be obtained from their original data repositories, as cited in the References. The Sri Lanka HIES 2019 dataset and the WFP VAM datasets are subject to access restrictions and are available from the original data providers upon reasonable request and subject to their data-use agreements.

%%%%%%%%%%%%%%%% END OF MAIN TEXT %%%%%%%%%%%%%%%

% Supplementary Information (compiled separately from supplement.tex)
\clearpage
\includepdf[pages=-]{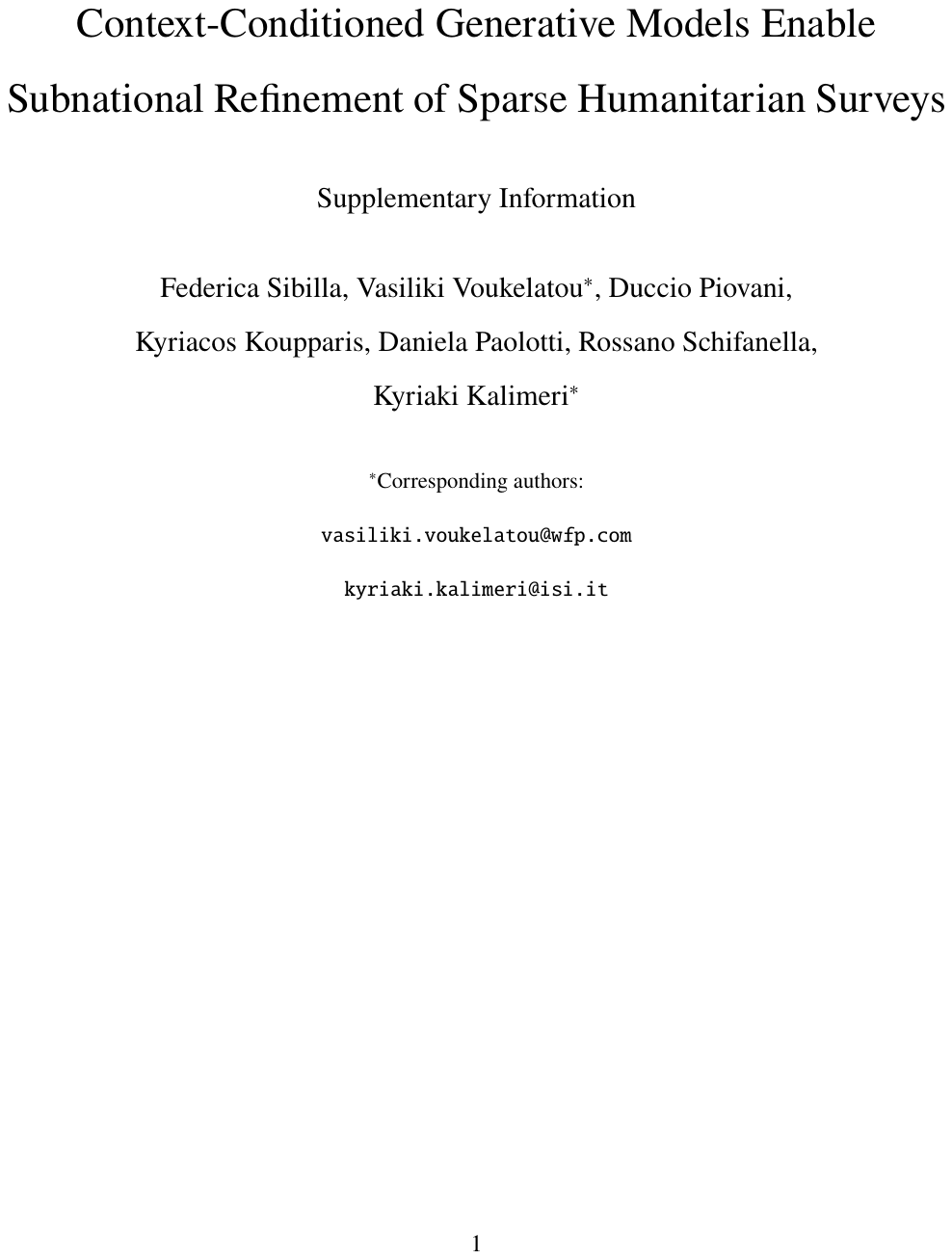}

\end{document}